\documentclass[10pt]{iopart}

\usepackage{graphicx}
\usepackage{amsfonts}
\usepackage{color}
\usepackage{hyperref}
\newcommand {\be}{\begin{equation}}
\newcommand {\ee}{\end{equation}}

\begin{document}

\paper[]{Large-deviations approach to thermalization: 
the case of harmonic chains with
conservative noise}
\date{\today}

\author{Stefano Lepri $^{1,2}$}
\address{$^{1}$  Consiglio Nazionale
delle Ricerche, Istituto dei Sistemi Complessi, 
via Madonna del Piano 10, I-50019 Sesto Fiorentino, Italy
}
\address{$^{2}$Istituto Nazionale di Fisica Nucleare, Sezione di Firenze,  
via G. Sansone 1 I-50019, Sesto Fiorentino, Italy}

\ead{stefano.lepri@isc.cnr.it \newline }

\begin{abstract}
We investigate the possibility of characterizing 
the different thermalization pathways through a large-deviation
approach. Specifically, we consider clean,  disordered and
quasi-periodic harmonic
chains under energy and momentum-conserving noise. 
For their associated master equations,  describing the dynamics
of normal modes energies, we compute the fluctuations of activity and 
dynamical entropy in the corresponding biased ensembles. 
First-order dynamical phase transition are 
found that originates from different activity
regions in action space. At the transitions, the 
steady-state in the biased ensembles changes from extended 
to localized, yielding a kind of condensation in 
normal-modes space.
For the disordered and quasi-periodic 
models, we argue that
the phase-diagram has a critical point 
at a finite value of the disorder or potential strength.

\end{abstract}

\section{Introduction}

Thermalization, the process by which a system reaches a state of thermal equilibrium with its surroundings, stands as one of the fundamental phenomena governing the behavior of diverse physical systems. From quantum particles in condensed matter systems to the dynamics of stars in astrophysical environments, the concept of thermalization underpins our understanding of equilibrium states and relaxation processes across various scales.
Particularly fascinating is the key role of relaxation timescales on the 
very possibility of observing signature of the past events and to 
store and retrieve memory of them \cite{rovelli2022memory}.  

On general grounds,  the problem is usually approached from two main sides. 
In the context of nonlinear dynamics,  emphasis is on how energy spreads
from a particular point or region in phase space towards the largest portion of 
it where energy equipartition among all degrees of freedom occurs. This is the 
familiar approach pioneered by Fermi, Pasta, Ulam and Tsingou, which has a long-standing 
history \cite{gallavotti2007fermi} but is still actively
investigated.  Closeness to integrability 
usually affects strongly the equilibration in a non-trivial way
\cite{Benettin2011,Benettin2013,DeRoeck2013,Huveneers2017,Fu2019,
goldfriend2019equilibration,baldovin2021statistical}.
The situation is even more intriguing for systems with quenched
disorder, where and interplay between Anderson localization 
and chaotic diffusion is at work
\cite{Pikovsky08,Kopidakis08,Skokos2009,Lepri2010,Basko2011}. 

The second  approach is more statistical, in the sense that it rests on some 
form of stochastic dynamics and on how relatively unlikely, low-entropy
states evolve towards maximal entropy one. This \textit{\'a la }Boltzmann approach
involves equations for distributions derived under suitable assumptions.
For instance, in the case of nonlinear lattices this leads to Boltzmann-Peierls \cite{spohn2006phonon}
and wave turbulence equations \cite{onorato2023wave} .
For one-dimensional chains this turned out to be useful to understand
anomalous transport  \cite{Pereverzev2003,Nickel07,Lukkarinen2008,lukkarinen2016kinetic} 
and the effect of  nonlinear resonances on (pre)thermalization \cite{Onorato2015,huveneers2020prethermalization}.

To  calculate the probability of these trajectories  and how  atypical ones
arise dynamically one can rely on  large deviation theory  \cite{touchette2009large,touchette2018introduction,jack2020ergodicity}.
In the last decades,  such approach has rapidly grown as an effective
tool to study fluctuations in
statistical mechanics and stochastic processes.  For dynamics,  
it provides a mean to compute the probability of untypical trajectories
and value of observables that depart  significantly from their average, typical ones. 
Dynamical phase transitions, namely non-analiticities of the 
large-deviation functions, may occur as a  
signal of qualitative changes in the 
statistical properties.

Our aim here is to explore the possibility to use this approach to explore 
the thermalization dynamics of many-body classical systems.  
The idea would be to quantify and characterize regions in phase space that
yield possibly slow, or at least sizeably different from average, relaxation pathways.
In principle, this should provide insight on the different phase-space 
trajectories followed during equilibration. 

A similar idea has been discussed in \cite{zannetti2014condensation}
(see also the account in \cite{corberi2019probability}). There it has been
shown that the large deviation of an observable may have singularities
even  in the context of the Gaussian model,  a paradigmatic non-interacting system, before and after an instantaneous temperature quench. The 
associated phenomenon of condensation of fluctuations is thus a characteristic
feature of the thermalization process. Remarkably, condensation
originates from the duality between large deviation events in the given system 
and typical events in a "biased" one.  
We also mention a conceptually related work on 
large deviations for wave-turbulence equations \cite{guioth2022path}.
We mostly have in mind the relaxation of isolated systems: 
as a matter of facts,  large-deviation approaches have been reported for
nonequilibrium steady states close to what we will deal here, 
like harmonic networks in contact with
thermal baths (see e.g \cite{saito2011generating,fogedby2012heat,jaksic2017entropic}
and references therein).

Since dealing with genuine non-linear forces is a formidable 
task, it is convenient to adopt an intermediate "mesoscopic" approach 
whereby nonintegrable interactions are replaced by a random noise.
This idea has proved to be useful,  for instance, 
to study the Lyapunov time-scale
 \cite{lam2014stochastic,goldfriend2023effective}
and  the onset of diffusive dynamics in almost-integrable systems 
\cite{lopez2023integrability}.  Also,  such approach
reproduces many features of nonlinear lattices in steady nonequilibrium
states \cite{basile2016thermal,Lepri10}.
This semplification may overlook some of the 
subtle correlations of the deterministic dynamics but can provide 
also quantitative results, as for instance the 
scaling law of
the Lyapunov exponent  with the energy density for the 
Fermi-Pasta-Ulam-Tsingou model \cite{goldfriend2023effective}.
In this spirit, we focus on 
a specific model: an harmonic network with momentum and
energy conserving noise that we studied in \cite{lepri2023thermalization}.
Such model has a the remarkable advantage that, under suitable 
approximations,   
it allows to write explicitly a linear master equation describing
energy spreading  in 
action space.  We will exploit this feature to compute the large 
deviation properties exactly, through standard methods.

The paper is organized as follows.  In Section \ref{sec:me} we present
our model system:  an harmonic network subject to 
momentum and energy conserving noise, which is the main object of study.
We first briefly recall  the kinetic approach presented in \cite{lepri2023thermalization}
that allows to write a master equation for diffusion in action space
which is the starting point of the subsequent analysis.
In Section \ref{sec:ld} we briefly recall the methods used to compute 
the cumulant generating function of the main observables, the dynamical
activity and the dynamical entropy.  The first results concerns 
the simplest case of a chain with translational invariance 
discussed in Section \ref{sec:tinv}. In the second part we examine 
the thermalization and large deviation properties of two paradigmatic
models, the disordered (Section \ref{sec:dis}) and quasi-periodic
chain (Section \ref{sec:aa}) that correspond, respectively,  to the famous 
Anderson and Aubry-Andr\'e -Harper models in the context of electronic 
transport.   

\section{Continuous-time master equation}
\label{sec:me}

Let us first consider a general master equation for the probabilities 
$P_\nu$ of a state, $\nu =1,\ldots,N$ with transition rates $R_{\nu,{\nu'}}$
\be
\dot P_\nu= \sum_{\nu'}\left[R_{\nu,{\nu'}} P_{\nu'}
- R_{{\nu'},\nu}P_{\nu}\right]. 
\label{releqp}
\ee
Eq. (\ref{releqp}) can be rewritten as
\be
\dot P_\nu= \sum_{\nu'\neq \nu}W_{\nu,{\nu'}} P_{\nu'}\qquad
W_{\nu,{\nu'}}=R_{{\nu'},\nu} - r_\nu \delta_{\nu,{\nu'}}  
\label{releqw}
\ee
where we introduced the escape rate from state $\nu$ as
\be
r_\nu = \sum_{\nu'\neq \nu}R_{{\nu'},\nu} 
\label{escaper}
\ee
and $1/r_\nu$ the mean residence time in state $\nu$. 
The calculation of the large-deviation properties for (\ref{releqp})
is well established and will be briefly reviewed in the next Section.

Before this,  we now specify the physical context that we will deal with. 
Let us consider the following quadratic Hamiltonian,        
\begin{equation} 
\label{ham_all2all}
H=\sum_{i=1}^{N} \frac{p_{i}^{2}}{2m_i}+\frac{1}{2}\sum_{i,j=1}^Nq_{i} \Phi_{ij}q_{j},
\end{equation}
where $(q_i,p_i)$, $i=1,\ldots N$ are the canonically-conjugated coordinates
ad momenta 
(for simplicity, we will deal henceforth with the equal-mass case,  and set $m_i=1$).
The coupling matrix $\Phi$ is semi-positive definite and symmetric so that 
its eigenvalues are real and non-negative. 
The equations of motion of the isolated 
network are 
\be
\ddot q_i = -\sum_j \Phi_{ij}q_{j}\quad .
\ee
In order to have momentum conservation it must be $\sum_i\Phi_{ij}=0$. 
In this general formulation, the model encompasses several different setups.
It includes the standard case of periodic lattices,  where
$\Phi$ is the familiar nearest-neighbor Laplacian matrix, but also 
the case of disordered structures with random and/or sparse connections 
as in the case of 
elastic networks \cite{bouchaud1990anomalous,hastings2003random}.
It can be regarded as an idealized model of, say, a large atomic cluster or 
a protein in its native state. 

The normal modes' coordinates $(\mathcal{Q}_\nu,\mathcal{P}_\nu)$ 
of the network have 
eigenfrequencies $\omega_\nu$ and are defined by the transformations 
\be
\mathcal{Q}_\nu=\sum_{l=1}^N q_l \,\chi_l^{\nu}, \quad  \mathcal{P}_\nu=\sum_{l=1}^N p_l \,\chi_l^{\nu}.
\label{modi}
\ee
The $\chi_l^{\nu}$, $\nu=1,\ldots,N$, are orthonormal and can be taken to be real
for the time being.  The main observables are the mode energies 
$\mathcal{E}_\nu=(\mathcal{P}_\nu^2+\omega_\nu^2\mathcal{Q}_\nu^2)/2$

The classic problem  amounts to ask how
a (possibly weak) nonlinear interaction potential added to  $H$ 
brings the harmonic modes to equipartition.
For conservative dynamics, upon expressing the Hamiltonian in the familiar
action-angle variables, a nonintegrable perturbation defines a network of 
couplings among the unperturbed actions  \cite{Mithun2018,danieli2019dynamical}. 
The connectivity of such a network will determine relaxation and ergodic properties.  
In general, one can distinguish such networks depending on whether  the number
of groups of actions linked by the perturbation scales
intensively or extensively on the number of degrees of freedom $N$. 
These are termed short or long-range networks, respectively, 
depending on whether the coupling range  if roughly constant or increases proportionally 
to $N$  \cite{Mithun2018,danieli2019dynamical}. 

In the nonlinear case the coupling may involve three or more actions, 
resulting in complicated higher-order interaction on a hyper-graph structure.
In view of the difficulty of the problem,  it is of interest to formulate a hybrid
dynamics for Hamiltonian (\ref{ham_all2all}) defined by adding 
on top of the linear deterministic dynamics a conservative stochastic process \cite{BBO06} defined as follows.  Consider the network state
at time $t$. At a later time $t+\tau$ there occur
a "collision" event whereby a couple of particles $(m,n)$, $m\neq n$
is randomly selected according to the joint probability 
$w_{n,m}$ and their momenta are exchanged, $(p_n,p_{m})\to (p_{m},p_n)$. 
This move conserves energy and momentum. 
Physically, it can be interpreted as a perfectly elastic rebound
as it would occur for a infinite square-well pair potential. 
The $w_{n,m}$ are usually assigned to describe local
interactions (e.g. nearest-neighbors) but may 
include long-range terms interaction across the lattice as well. 
The intervals between subsequent collision times $\tau$ are also 
taken as random variables with some preassigned distribution 
with given, finite,  average $\langle \tau \rangle$.

In the kinetic approximation \cite{lepri2023thermalization}
the (angle-averaged) energy $E_\nu=\overline{\mathcal{E}_\nu}$ of each normal mode obeys the equation 
\be
\dot E_\nu= \sum_{\nu'\neq \nu}\left[R_{\nu',{\nu}} E_{\nu'}
- R_{{\nu'},\nu}E_{\nu}\right] 
\label{releq}
\ee
This can be regarded as an averaged collision operator
coarse-grained over a time scale which is longer than the typical collision
time.  We work with finite
$\gamma=\frac{1}{N\langle \tau \rangle}$,  that corresponds to choosing a finite collision probability per site (as usually employed in the simulations \cite{Lepri10,Delfini10,Iacobucci2010,lepri2020too}). 
The form of equation (\ref{releq})  guarantees that $\sum_\nu E_\nu$ remains constant
as it should. One major advantage is that, 
the simple form of the collision rule allows us to compute explicitly
the transition rates in terms of the eigenvectors \cite{lepri2023thermalization}.    First, define $V$ as
the a unit vector of components 
\begin{equation}
V_\nu=V_\nu^{(n,m)} \equiv \frac{\chi_n^{\nu} -\chi_m^{\nu}}{\sqrt{2}},
\end{equation}
Note that $V=V^{(n,m)}$ is a random vector as the indexes $n,m$ are chosen
at random with the prescribed rule.  
We refer the reader to \cite{lepri2023thermalization} for the 
details and just mention the formulas needed henceforth.
It is found that
\begin{equation}
R_{\nu,{\nu'}} = \frac{2}{\langle \tau \rangle} \sum_{(n,m)}
(V_\nu^{(n,m)})^2 \,w_{n,m} (V_{\nu'}^{(n,m)})^2 \equiv \frac{2}{\langle \tau \rangle}\overline{V_\nu ^2V_{\nu'}^2}
\label{rates}
\end{equation}
where the overline is a shorthand notation  for the average over 
$w_{n,m}$ of the random collisions. Note that $R_{\nu,{\nu'}}=R_{\nu',{\nu}}$
with the escape rate from state $\nu$ as
\be
r_\nu = \sum_{\nu'\neq \nu}R_{{\nu'},\nu} = \frac{2}{\langle \tau \rangle}\overline{(1-V_\nu ^2)V_{\nu}^2}.
\ee
For a generic coupling  matrix $\Phi$, such that there 
is no decoupling among different subsets they 
are all nonvanishing. 
They are of the form of the well-known Fermi golden rule, involving
the squared amplitudes of the eigenmodes.

Formally, the quantities $ P_\nu=E_\nu/\sum E_\nu$ (comprised between 0 and 1)
satisfy (\ref{releq}).
A physical  interpretation would be to imagine an large ensemble of quasi-particles
with a given distribution among modes, each $P_\nu$ being the instantaneous 
fraction of them in mode $\nu$.  
Altogether, the dynamics is a random walk of quasiparticles in action 
space. Transitions  
are determined by the eigenstate structure of the coupling matrix $\Phi$, 
and not assigned a-priori. 


Following the usual arguments of Markov processes, the system is 
ergodic and approaches microcanonical equilibrium
where $P_\nu=1/N$ (equipartition) according to the 
properties of the "collision operator" defined by (\ref{releqp})
and (\ref{rates}).
So the thermalization problem reduces to computing its 
$N$ eigenvalues $\mu_\nu$,  $\nu=1,\ldots,N$. 
whose absolute values give the  spectrum of relaxation rates.   The first 
eigenvalue is $\mu_1=0$ and corresponds, as said,  to the steady state of energy 
equipartition. On physical grounds, $\mu_1$ must be non-degenerate 
since we expect the dynamics to be ergodic under the above hypotheses.
All the others $\mu_\nu$ must be strictly negative \cite{schnakenberg1976network}.
In particular,  the long-time relaxation rate is controlled by spectral 
gap $|\mu_2-\mu_1|=|\mu_2|$, $\mu_2$ being referred to as Fiedler eigenvalue in the context of diffusion on
graphs.
For a finite network, the scaling with $N$ of 
$\mu_2$ can be expressed in terms of the scaling of the spectral density.
More precisely,  
denoting by $\rho(\mu)$ the integrated
(cumulative) 
density of eigenvalues (i.e. the fraction of eigenvalues 
less than $\mu$), if
\be
\rho(\mu) \sim |\mu|^{d_\mu/2}, \qquad \mu \to 0 
\label{ds}
\ee
then $d_\mu$ is the spectral dimension of the action network.  We 
emphasize that this is a distinct quantity with the standard spectral dimension
of the Laplacian matrix $\Phi$ entering in the spectral 
problem for the Hamiltonian (\ref{ham_all2all}).

The solution of master equation with initial condition $P_\nu(0)$, 
is fairly straightforward.  Upon expanding 
in the basis of the eigenvectors $\psi^{(\nu)}$ of the 
matrix $W$ (the stochastic matrix 
is symmetric so there is no need to distinguish left and right eigenvectors), 
one can express the solution as
\begin{equation}
 P_\mu(t)=\sum_{\nu\nu'} e^{\mu_\nu t}  \psi^{(\nu)}_\mu  \psi^{(\nu)}_{\nu'} P_{\nu'}(0).
\end{equation}
This can be easily solved numerically to compute for the 
observable of interest.

\section{Large-deviations approach}
\label{sec:ld}

In this section we briefly review the methods employed in the following.
The material is standard in the related literature, and we mostly
follow reference \cite{lecomte2007thermodynamic}.

\subsection*{Activity}

One quantity studied in the literature is the activity,
which counts the number of jumps in a trajectory
i.e. the number $K$ of  configuration changes during a time $t$. 
As the name implies, it is a measure of how active a given trajectory is:
an active trajectory has many changes of configuration, an
inactive one just few or none.

The relevant operator to study is the tilted generator
which for the activity is defined by the $N\times N$ matrix  with elements \cite{garrahan2007dynamical,garrahan2009first} 
\begin{equation}
\left(W_K(s)\right)_{{\nu'},\nu}\equiv e^{-s}R_{{\nu'},\nu}  - r_\nu \,\delta_{\nu,{\nu'}}.  
\end{equation}
Note that his operator has  $-r_\nu$ on the diagonal and the 
first terms appears only in off-diagonal entries 
(see for example eqs.  (44) and (71) in Ref. \cite{lecomte2007thermodynamic}).
  
Following the general strategy,  the large-deviation  statistics is 
encoded in the leading eigenvalue
and (right and left) eigenvectors of the tilted operator
(also called scaled cumulant generating function). More precisely, we denote 
by  $\lambda_K(s,N)$ the largest eigenvalue of 
$W_K(s)$ along with its the right eigenvector defined by
$W_K(s)\phi_K=\lambda_K\phi_K$. Upon normalization, 
$\phi_K$ can be considered as a steady  probability distribution for the "biased"
dynamics (sometimes  termed $s$-ensemble).
As it is known \cite{lecomte2007thermodynamic}
for  $s = 0$ the eigenvector yields the density in the steady state, which
in our case is the equipartition 
which is homogeneous.  For $s < 0$, it probed
the regime in which the mean activity $K /t$ of histories is typically larger than
in the steady state. They correspond to explored configurations where the density
is larger than the steady state distribution. On the other hand, at $s > 0$ histories
with smaller $K /t$ are favored. 

As an order parameter we may thus consider (see eq. (177) in \cite{lecomte2007thermodynamic}). 
\begin{equation}
\rho_K(s) = \frac{1}{N}\sum_{\nu=1}^N \nu {\phi_{K,\nu}}(s)
\label{ordpar}
\end{equation}
that, in our context, measure the occupancy of the steady state and thus
of the distribution of modes that are involved
in the biased dynamics.

\subsection*{Entropy production}

Another observable of interest  is associated to 
dynamical complexity, $Q_+=\ln Prob(history)$.
In the dynamical systems  literature  the corresponding 
large-deviation function is known as the topological pressure.
It can be computed in a similar
way through the operator  $W_+(s)$ (see e.g. Eq. (36) in Ref.
 \cite{lecomte2007thermodynamic})
\begin{equation}
\left(W_+(s)\right)_{{\nu'},\nu} \equiv R_{{\nu'},\nu}^{1-s}\,  r_{\nu'}^s  - r_\nu \,\delta_{\nu,{\nu'}}  
\end{equation}
and its  the maximal eigenvalue is the topological pressure, large deviation
of the entropy production. Following the same prescription as above, we 
seek for the leading eigenvalue $\lambda_+(s,N)$ with right eigenvector
$\phi_+$,  $W_+\phi_+=\lambda_+\phi_+$.  An order parameter 
$\rho_+(s)$ can
be defined as in (\ref{ordpar}) with $\phi_{+,\nu}$ replacing $\phi_{K,\nu}$.
The average $Q_+/t$  is the
Kolmogorov-Sinai entropy $h_{KS}= d\lambda_+/ds(0)$; another
important quantity is the topological entropy $h_{top}=\lambda_+(1)$ which
is the growth rate of the number of possible histories with time.

\section{Translationally-invariant chains}
\label{sec:tinv}

Let us start from
one-dimensional chains with  
translational invariance i.e. the 
case where $\Phi$ is a circulant matrix. The simplest example
would be the standard nearest-neighbor coupling, but the 
results apply to any circulant $\Phi$, including the case of
harmonic
long-range interactions \cite{andreucci2023nonequilibrium}.
In fact, in all those instances, the eigenvectors
are the familiar lattice Fourier modes and the vector $V$ can
be given explicitly
\[
\chi_l^\nu = \frac{1}{\sqrt{N}}e^{ik_\nu l}; \quad
V_ \nu^{(n,m)} \equiv \frac{ e^{ik_\nu n} -e^{ik_\nu m}}{\sqrt{2N}},
\]
where $k_\nu=\frac{2\pi \nu}{N}$ (we use the labeling $\nu=-N/2+1 \ldots N/2$ 
throughout
this Section). Thus 
the transition rates in formula (\ref{rates}) are known exactly once
$w_{n,m}$ is given. Furthermore, for any choice of $\Phi$,
the rates only depend on $w_{n,m}$. 
For definiteness, we will focus on the case of
nearest-neighbor random exchange, namely $m=n+1$. This is the 
mostly studied choice in the literature both for short (nearest-neighbor) 
\cite{BBO06,basile2010energy} than for long-range interactions \cite{tamaki2020energy}.
In this case,   the transition rates can be computed explicitly 
since \cite{lepri2023thermalization}
$$
|V_\nu^{(n,m)}|^2= \frac{1}{2N}|1-e^{ik_\nu}|^2 = \frac{2}{N}\sin^2\frac{k_\nu}{2},
$$
so that
 \begin{equation}
R_{\nu,\nu'}=
\frac{8\gamma}{N} \sin^2\frac{k_\nu}{2}\sin^2\frac{k_{\nu'}}{2}
\label{Rnn}
\end{equation}
\be
r_\nu = \sum_{\nu'\neq \nu}R_{{\nu'},\nu} =
4\gamma \sin^2\frac{k_\nu}{2} 
[1-\frac{2}{N}\sin^2\frac{k_\nu}{2}]
\ee
Note that the escape rate $r_\nu$ is different for each state $\nu$.
Another interesting property is that the off-diagonal terms are of order $1/N$, and thus the 
eigenvalues of the matrix $W$ are well approximated by the diagonal
elements 
\be
\mu_\nu \approx -4\gamma\sin^2\left(\frac{\pi (\nu-1)}{N}\right),
\label{mu} 
\ee
and the $\nu$th eigenvector is localized on $\pm \nu$ with all other 
components being small of order $1/N$. The relaxation rate of 
a generic non-equilibrium initial condition occurs, at long enough times, 
with a rate $|\mu_2|\approx\gamma (\pi/N)^2$.
Thus,  the matrices $W_K$ and $W_+$ are also known analytically:
for instance the off-diagonal elements of $W_+$ are 
\be
(W_+)_{\nu'\neq \nu}= R_{{\nu'},\nu} ^{1-s}\,  r_{\nu'}^s
= \gamma \frac{2^{2s+3}}{N^{1-s}} \sin^2\frac{k_\nu}{2} 
\left(\sin^{2}\frac{k_{\nu'}}{2}\right)^{1-s}.
\ee

\begin{figure}
\centering\includegraphics[width=0.7\textwidth]{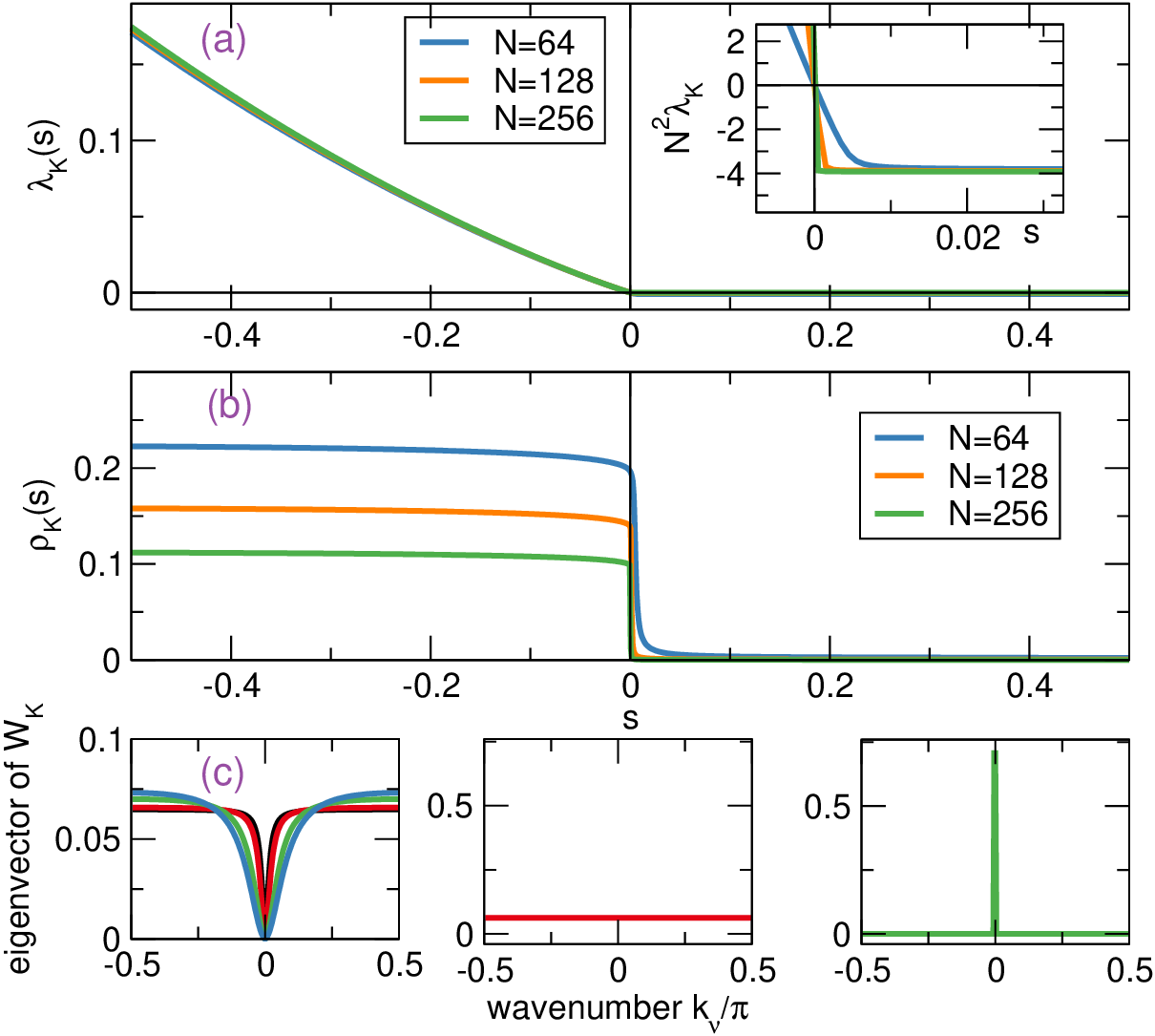}
\caption{Large deviations of activity for the translationally invariant chain
with conservative noise, $\gamma=1$. (a) the largest eigenvalue of $W_K(s)$ as 
a function of $s$,  for different chain lengths $N$; the inset in an enlargement
around $s=0$ showing the eigenvalues multiplied by $N^2$;
(b) the order parameter $\rho_K(s)$; (c) 
the eigenvectors $\phi_K$ for $N=256$ as a function
of the wavenumber $k_\nu$: 
left panel $s-0.005,-0.01,-0.0500.1$, middle $s=0$, right $s=0.1$.
The data show evidence of a dynamical phase 
transition at $s=0$ at which the activity displays
a first-order jump and  $\phi_K$
changes from extended to localized.
}
\label{fig:activ}
\end{figure}

Let us first discuss the case of the activity,
In Figure \ref{fig:activ}a we report the numerically computed
$\lambda_K(s)$ for different sizes.
It is shown that that large deviation function has 
a discontinuity in the derivative at $s=0$ corresponding to a dynamical 
first-order phase transition. Indeed, 
increasing $s$ leads to a sudden jump in the typical
activity $d\lambda_K/ds$, which corresponds to a dramatic change in the 
number of configurations
explored by histories with  activity rate $K /t$ (see the inset
of Figure \ref{fig:activ}a) .
As far as the system size scaling is concerned, for $s<0$ $\lambda_K$ is 
$N-$ independent, 
while for  positive $s$ it vanishes as $1/N^2$ (see inset in Fig. \ref{fig:activ}a).
This finite-size analysis provide a convincing evidence that the discontinuity
persists in the thermodynamics limit.

\begin{figure}
\centering\includegraphics[width=0.7\textwidth]{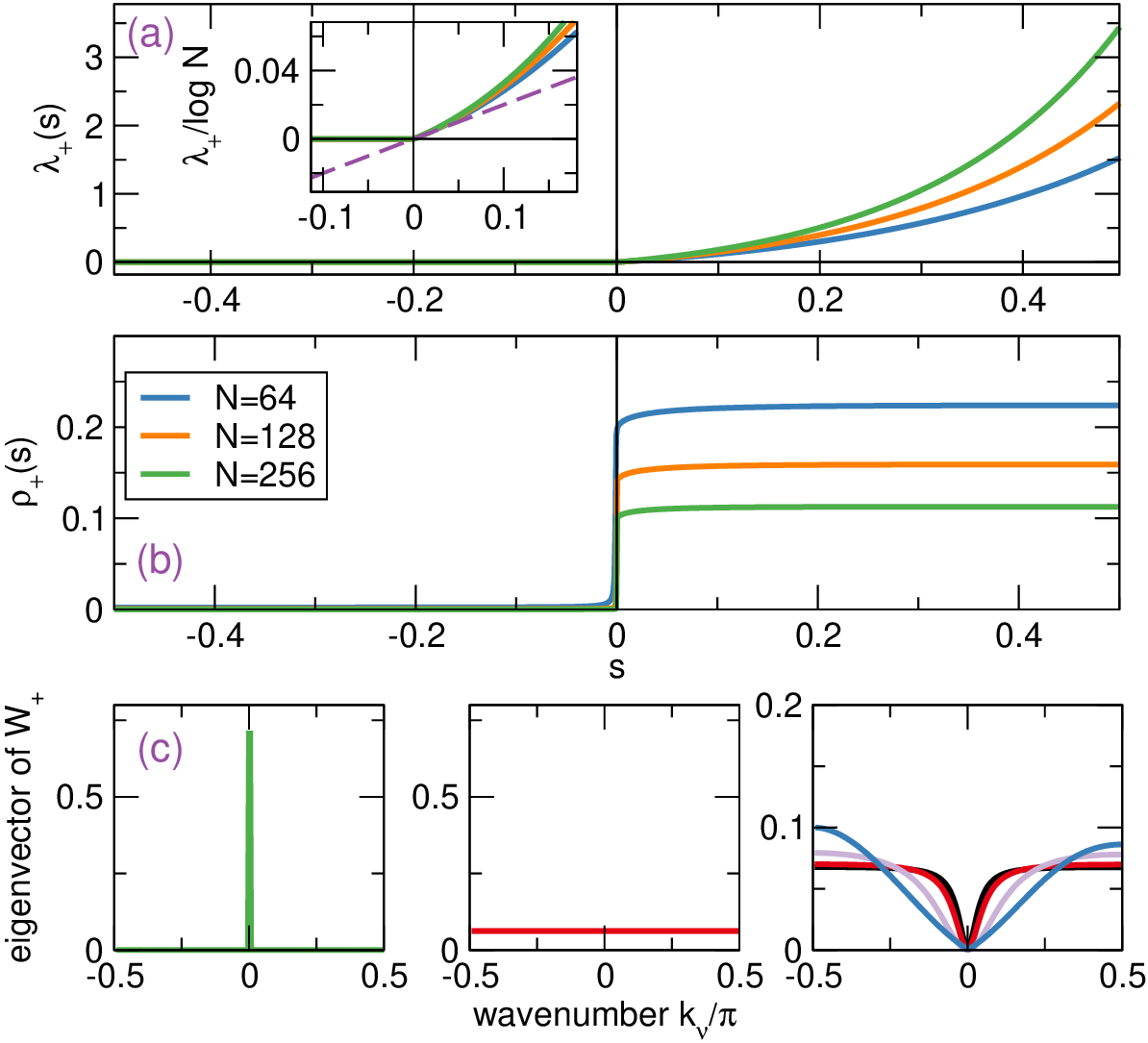}
\caption{Large deviation function for the dynamical entropy for the 
translationally invariant chain
with conservative noise, $\gamma=1$. (a) The largest eigenvalue of $W_+(s)$ as 
a function of $s$  for different chain lengths $N$; the inset in an enlargement
around $s=0$ showing the eigenvalues divided by $\ln N$ and the
dashed line has a slope $2\gamma$;
(b) the parameter $\rho_+(s)$; (c) the eigenvectors $\phi_+$ 
for $N=256$ as a function
of the wavenumber $k_\nu$: 
left panel $s=-0.1$, middle $s=0$, right $s=0.005,0.01,0.050,0.1$ .}
\label{fig:tent}
\end{figure}

Taking into account the fact that plane waves come in pairs with opposite 
wavenumber,  it is appropriate to adapt the definition of the 
order parameter (\ref{ordpar}) as
\begin{equation}
\rho_K(s) = \frac{1}{N}\sum_\nu |k_\nu| \phi_{K,\nu}(s)
\end{equation}
(and the same for $\rho_+(s)$).
This gives indication about the typical wavenumbers which are 
mostly active along a biased trajectory.
Figure \ref{fig:activ}b shows that $\rho_K$ has a discontinuous jump
at the transition point $s=0$, indicating that different 
subset of degrees of freedom are involved in the biased dynamics.
This is confirmed by inspecting the structure of the leading 
eigenvector $\phi_K$, which is markedly different it the two
regimes, see  Figures \ref{fig:activ}c:
\begin{itemize}
\item at $s=0$ the eigenvector $\phi_K$ is uniform, as expected since
it must correspond to the steady state of energy equipartition;
\item  for $s>0$, $\phi_K$ has two large components at $\pm k_1$ plus 
a uniform 
background of size $O(1/N)$; moreover, this shape it is independent of $s$;
\item for $s<0$, $\phi_K$ is very small at the band center $k_\nu=0$
were it attains a minimum
and increasingly concentrates toward 
zone-boundary Fourier modes $k_\nu\approx \pm \pi$
upon decreasing $s$.
\end{itemize}
Thus, trajectories in the inactive phase $s>0$ consists mostly 
the  longest-wavelength mode which remains excited, 
undergoing very little exchange with others.
On the other hand, the active phase $s<0$ is mostly peaked 
on short-wavelengths and involve rapid energy exchange among 
a finite fraction of them.  
This gives an overall picture of
action space, which is consistent with the numerical 
simulation of the microscopic dynamics \cite{lepri2023thermalization}.  
The remarkable fact that the transition occurs at $s=0$ 
means that the model dynamics displays "phase coexistence" 
between fast and slow relaxing modes. 

To complement the analysis, we also performed a calculation of the 
large deviation of the dynamical entropy, 
Figure \ref{fig:tent} reports the same type of analysis  as above.  
There is still a similar transition at $s=0$,  separating
two regimes: one "regular" with zero KS entropy and one
"chaotic" with finite (perhaps $N$ dependent) entropy.
So the natural dynamics present a coexistence of these two.
Since the standard KS entropy $h_{KS}$ is given by the the slope at the origin, 
from the data in the inset  of Figure \ref{fig:tent}a it is seen that
for $s\to 0^+$,  
$h_{KS}\approx2\gamma \log N$, meaning that it subextensive 
in the system size. 
It is worth noticing that such scaling occurs in other instances,  like 
for instance the infinite-range Ising model (see Section 6 in \cite{lecomte2007thermodynamic}).
This may be related to the fact that the coupling among modes 
is of mean-field type, see again eqs. (\ref{Rnn}). 
This feature is different from genuinely 
chaotic dynamical systems,  where typically the KS entropy
is extensive, $h_{KS} \sim N$.  

At this point, we remark that there is  indeed a 
similarity with  
kinetically constrained models,
as Fredrickson-Andersen dynamics and the like, 
which have been thoroughly studied as toy models of glasses 
\cite{garrahan2007dynamical,jack2020ergodicity}. Also there, 
the  large-deviation function is singular at $s=0$ and the order
parameter has a first-order jump. The interpretation is
that there are two
phases, an active one for $s < 0$ and an inactive one for $s > 0$.
Physical dynamics take place at $s=0$, where the two dynamic
phases coexist.  It  our case the same interpretation applies in 
action rather than in physical space.

\section{Disordered chain}
\label{sec:dis}

Let us now turn to the simplest non-homogeneous lattice:
the harmonic chain with disorder in the pinning potential
\be 
H = \sum_{i=1}^N \left[{p_i\over 2}^2
+\frac12 (1+\sigma_i) q_i^2 +
\frac{1}{2}(q_{i+1}-q_{i})^2\right] 
\label{dishami}
\ee
with $\sigma_i$ being i.i.d. variables with 
uniform distributions in $[0,w]$,  gauged by the disorder 
strength parameter $w$. Periodic boundary conditions are assumed.
As it is well known, the eigenstates are the exponentially-localized 
Anderson modes, whose localization length decreases with $w$ \cite{Matsuda70,Visscher71}. 
Several variants in this type of model with conservative noise
have been considered earlier
\cite{bernardin2008thermal,dhar2011heat,bernardin2019hydrodynamic}.
We consider the standard case of nearest-neighbor collisions
and individual realizations of the quenched disorder $\{\sigma_i\}$.

\begin{figure}
\hfill\includegraphics[width=0.85\textwidth]{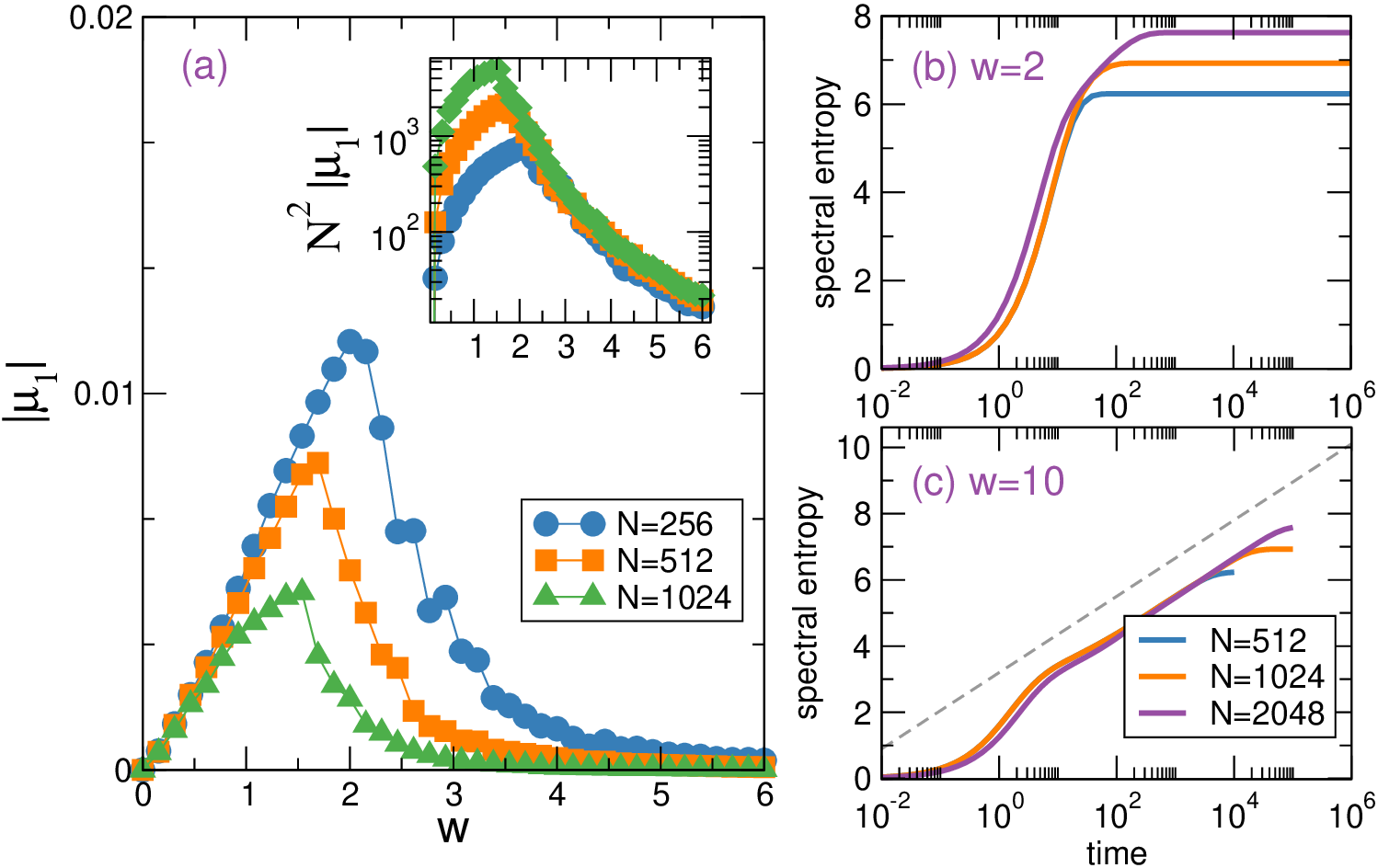}
\caption{Thermalization of the disordered chain with conservative noise
$\gamma=1$:
(a) the spectral gap as a function of the disorder strength $w$,  the inset
shows that $\mu_2$ is proportional to $1/N^2$ for $w$ large; (b,c)
time evolution of the spectral entropy $S(t)$ for two values of $w$ 
corresponding to weak and strong disorder respectively.  Data from 
numerical solution of the master equation starting from the ground-state mode $\nu=1$.The dashed line is the law $S(t) = \frac{1}{2}\log t + const.$.}
\label{fig:gapdis}
\end{figure}

Before discussing large deviations, we illustrate some features that 
can be inferred from the kinetic equations (see also \cite{lepri2023thermalization}
for a comparison with numerics). 
The localization level affect drastically the connectivity
in action space as measured by the transition rates. Indeed, the $R_{\nu,\nu'}$ 
are all non-vanishing, but can be exponentially small due to the 
small overlap of the corresponding eigenvectors, see again
Eq. (\ref{Rnn} ).
In Figure \ref{fig:gapdis}a
we plot the spectral gap as a function of disorder. There is 
a size-dependent disorder value, separating  fast relaxation
regime from a slower one. Qualitatively, this correspond to a change 
in the structure of the matrix $W$ from fully connected (as in the 
ordered case) to almost-banded. Above such value, the spectral
gap vanishes as $1/N^2$, see the inset in  Figure \ref{fig:gapdis}a.

A key indicator that has been often employed 
to characterize thermalization is the 
non-equilibrium (Shannon) spectral entropy 
\cite{livi1985equipartition}
\be
S(t) = -\sum_\nu P_\nu \log P_\nu.
\label{entropy}
\ee
For large $t$, $S$ approaches its equipartition value $\log N$. In general
$S$ will depend on the initial conditions.  In Figure \ref{fig:gapdis}b,c
we report the evolution of $S(t)$ for two values of the disorder strength
and different sizes obtained by numerical solution 
of the master equation, starting from the ground-state excitation 
$P_\nu(0) = \delta_{\nu,\nu_0}$. ($\nu_0=1$ in the calculations). There is a crossover from an initial exponential 
to an intermediate logarithmic growth. The initial regime can be accounted 
for by a  "mean-field" approximation  
\cite{mulken2017information}: assuming that energy of the initially excited mode 
is evenly redistributed among the other $N-1$ ones, namely $E_\nu\approx (1-E_{\nu_0})/(N-1)$ for $\nu\neq{\nu_0} $
yielding the approximation
\be
S(t) \approx  S_{MF}\equiv -E_{\nu_0} \log E_{\nu_0}-(1-E_{\nu_0})
\log\left(\frac{1-E_{\nu_0}}{N-1}\right)
\label{appentropy}
\ee
where $E_{\nu_0}(t)\approx (1-1/N)\exp(-\mu_{\nu_0} t)+1/N$. 
This approximation works well for long-range action networks i.e.
for weak disorder.
The second regime
is readily understood as a one-dimensional diffusive process,  where the number
of excited modes grows as a $\sqrt{t}$.   This is in agreement
with the  known result
that for diffusion on graphs
the Shannon entropy grows as $\frac{d_s}{2}\log t$,  
if the spectrum of the corresponding Laplacian operator 
shows scaling with a finite spectral dimension $d_s$
\cite{mulken2017information}. 
It also fits with the observed wavepacket spreading law  in
nonlinear disordered chains in the strongly chaotic regime 
\cite{Skokos2009,flach2010spreading} where indeed microscopic chaos
can be effectively approximated by a stochastic process.

\begin{figure}
\hfill\includegraphics[width=0.75\textwidth]{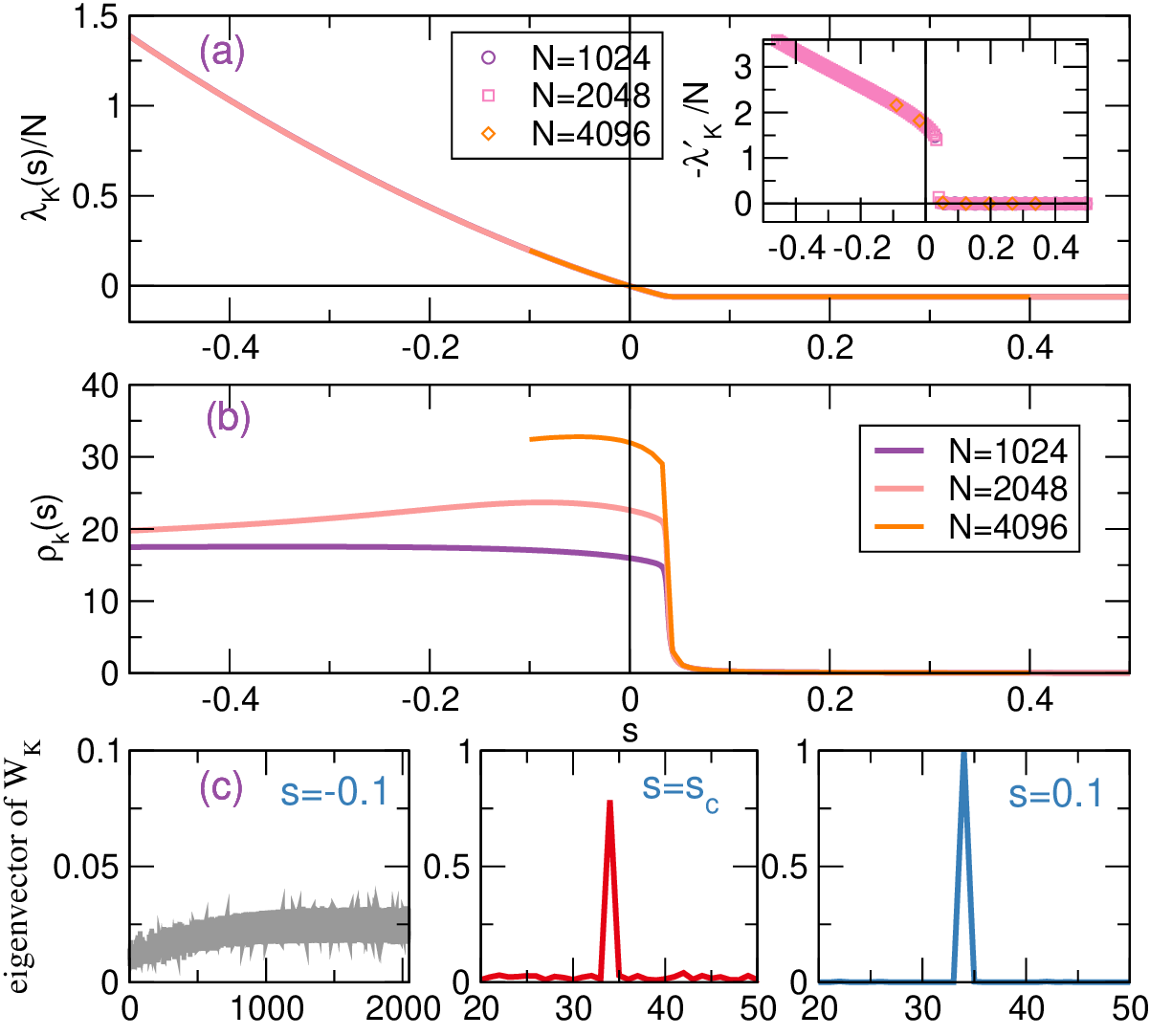}
\caption{ Large deviations of activity for the disordered chain 
with conservative noise $w=2$, $\gamma=1$: 
(a) the rescaled largest eigenvalue of $W_K(s)$ as 
a function of $s$  for different chain lengths $N$; 
In the inset we report the 
negative of the numerical derivative $-\lambda'_K(s)/N$ 
computed by finite differences.
(b) the order parameter $\rho_K(s)$; (c) the eigenvectors 
$\phi_K$ for $N=2048$. The data 
show evidence of a dynamical phase 
transition at $s=s_c\approx 0.0376$ at which the activity displays
a first-order jump and the leading eigenvector of $W_K$
changes from extended to localized.}
\label{fig:activdis}
\end{figure}

Let us now turn to the calculation of the large deviation properties. 
In Figure \ref{fig:activdis}a we  report data for the activity.
There is again a signature of a first-order dynamical transition as 
in the translationally-invariant case and thus of an active and an 
inactive phase.  However,  the critical point is now shifted to 
a nonzero value  $s=s_c$.  As for the finite-size analysis, note 
that the curve collapse upon dividing by $N$, signaling extensivity,  
at variance with the ordered case. Moreover, 
the data in  Figure \ref{fig:activdis}a show that for 
$N> 10^3$ the curves collapse very accurately also close to
$s_c$. Both the activity $d\lambda_K(s)/ds$ (inset of Figure \ref{fig:activdis}a) than 
the order parameter $\rho_K(s)$ (Figure \ref{fig:activdis}b) display the same qualitative behavior, 
with a step discontinuity at $s=s_c$. Moreover, we observed that $s_c$ increase upon
increasing the disorder strength $w$.  
The shift of the critical point implies that the unbiased dynamics
at $s=0$ has a finite activity rate.  
The structure of the eigenvector  $\phi_K$ displays a qualitative change 
at $s=s_c$ from a rather homogeneous profile to a localized one
(Figure \ref{fig:activdis}c). 
 
\begin{figure}
\hfill\includegraphics[width=0.75\textwidth]{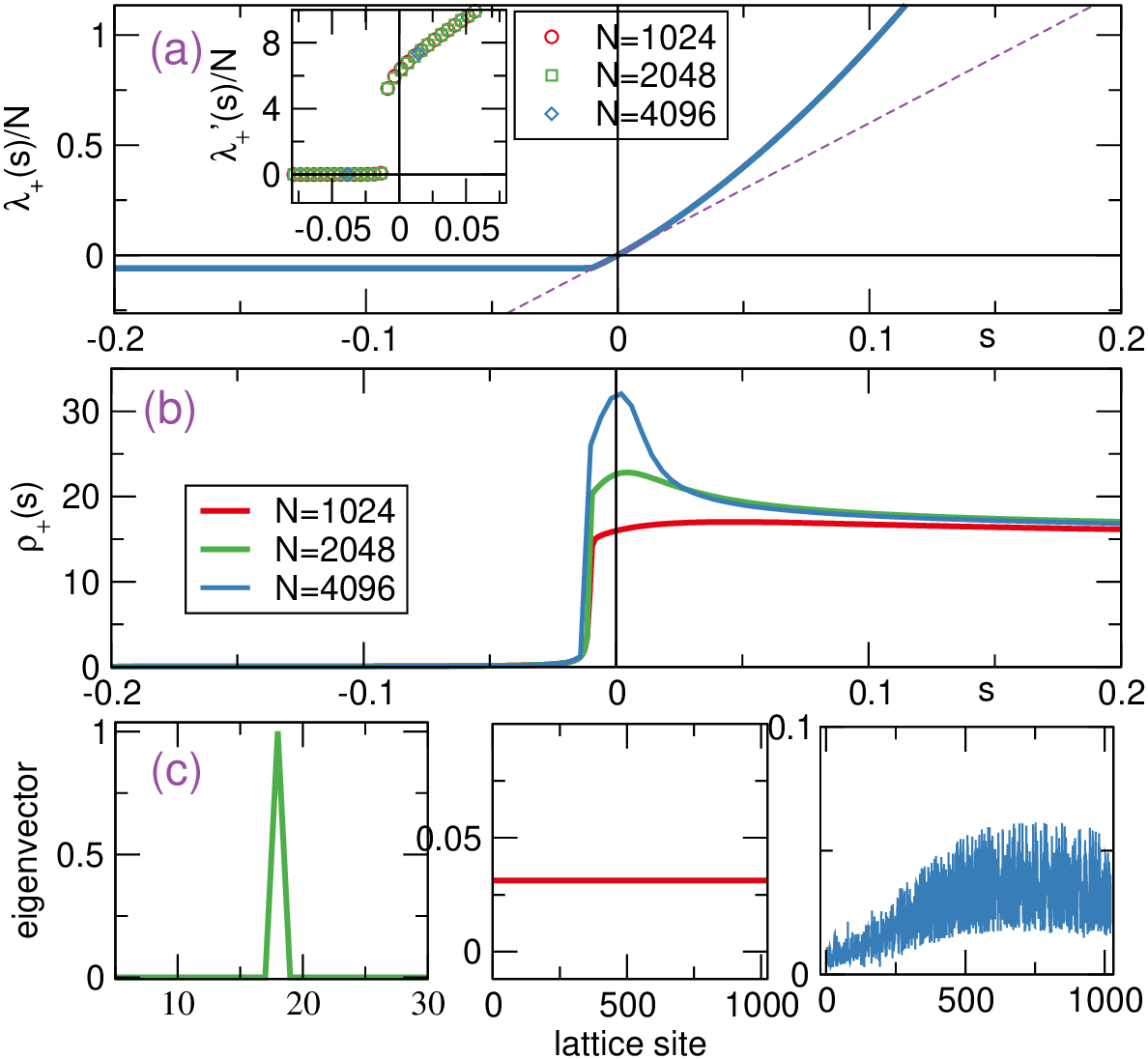}
\caption{Large deviations of entropy production for the disordered chain
with conservative noise $w=2$ $\gamma=1$: (a) the largest eigenvalue of $W_+(s)$ as 
a function of $s$  for different chain lengths $N$;  
the dashed line has a slope $2\gamma$.In the inset we report the 
the numerical derivative $\lambda'_K(s)/N$ 
computed by finite differences.
(b) the parameter $\rho_+(s)$; (c) the eigenvectors for $N=1024$ as a function
of the lattice site number left $s=-0.1$, middle $s=0$, right $s=0.1$
evidence of transition at $s\approx -0.059$ at which the eigenvector
changes from extended to localized.}
\label{fig:topdis}
\end{figure}

In Figure \ref{fig:topdis}, we report the calculation for the entropy. 
The data confirms that the dynamical transition is shifted to a finite 
$s$ where $d\lambda_+(s)/ds$ and the order parameter $\rho_+(s)$ undergo a finite jump
for large enough $N$.  A noteworthy difference with the respect to the 
ordered case is that the eigenvalue is proportional to 
$N$ in the whole $s$ range. Thus also the slope at the origin is proportional to $N$, meaning that in this case there is a finite 
Kolmogorov-Sinai entropy density $h_{KS}$.  This feature suggest a form of
more homogeneous "chaos" in presence of heterogeneity.  Finally, 
the shape of the eigenvector  $\phi_+$ displays a qualitative change 
at $s=s_c$ from a rather homogeneous profile to a localized one
(Figure \ref{fig:activdis}c), similar to what observed for the activity. 

What is the dynamical phase diagram of the model? 
We have seen that the phase transition persist upon increasing 
$w$.  On the other hand, for large disorder
the analysis in \cite{lepri2023thermalization} and 
the data in Figure \ref{fig:gapdis}c suggest that, for fixed $N$, 
the thermalization 
basically amounts to a diffusive process on a network with finite spectral dimension
(equal to one).
In this limit,  the problem should thus amount to a continuous-time random
walk and we do not expect any dynamical phase transition.  Indeed,  
for a  such a process $\lambda_{+} (s)$ should be roughly 
Poissonian  with no singularity \cite{lecomte2007thermodynamic}.

\begin{figure}
\hfill\includegraphics[width=0.8\textwidth,clip]{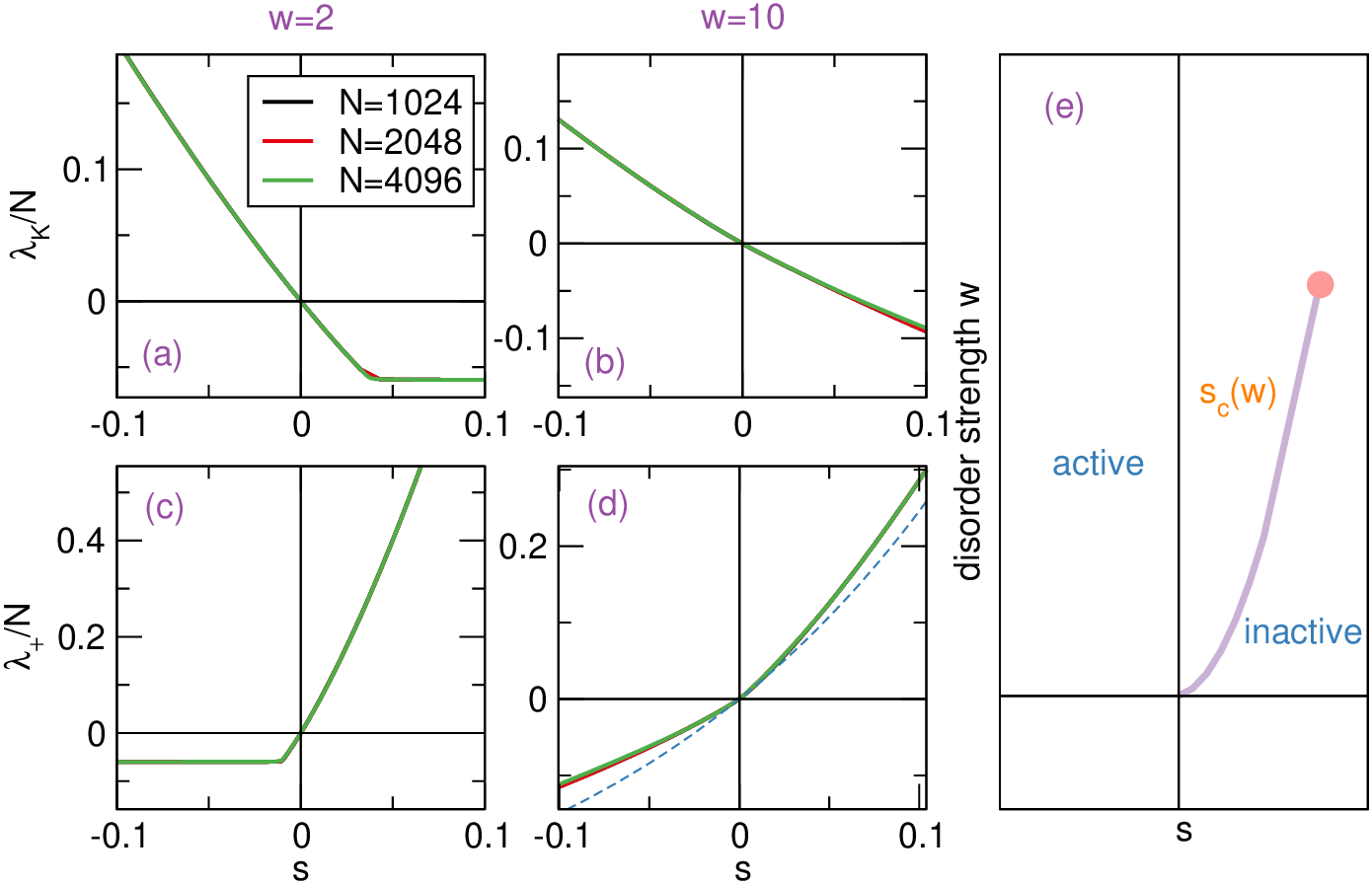}
\caption{Disordered chain with conservative noise: comparison of the 
scaled large deviation functions for two values of the disorder strength $w=2,10$
(left and middle panels respectively). (a,b) scaled largest eigenvalue
$\lambda_K$ and (c,d) $\lambda_+$. 
The dashed line in panel (d) is a fit with a Poissonian law yielding $0.38 (e^{5.0 s}-1)$; (e) schematic phase diagram  for the 
activity in the $(s,w)$ plane
where the continuous line is the transition line from active to inactive 
state. The red dot marks the critical point where the transition disappear.
}
\label{fig:pdiag}
\end{figure}

Based on the above limiting cases, we may argue that the transition
line $s_c(w)$ in the $(s,w)$ plane should terminate at a critical point.
In Figure \ref{fig:pdiag}a-d we show $\lambda_K(s)$ and $\lambda_+(s)$
divided by $N$ for two different values of the disorder strength.
The data confirm the expectation: for $w=10$ the curves have 
smooth derivatives and there is no sign of criticality.
In Figure \ref{fig:pdiag}e we draw a schematic phase diagram for the 
activity based on the above consideration. A similar diagram should be 
valid for the entropy.

\section{Quasi-periodic chain}
\label{sec:aa}

To further test the approach and the above results, 
let us consider the chain with a quasi-periodic on-site potential
\be 
H = \sum_{i=1}^N \left[{p_i\over 2}^2
+\frac12 \left[1+\Delta \cos(2\pi\beta i +\varphi) \right] q_i^2 +
\frac{1}{2}(q_{i+1}-q_{i})^2\right] .
\label{aa}
\ee
Here,  $\Delta>0$ is strength of the local force  and  
the parameter $\beta$ measures the degree of
incommensurability of the potential with the lattice period and 
$\varphi$ in an arbitrary phase. 
The eigenstate problem can be recast in the form 
\begin{equation}
\epsilon \chi_l = \chi_{l+1}+ \chi_{l-1} + 
\Delta \cos(2\pi\beta l +\varphi)\,\chi_{l}
\end{equation}
(where $\epsilon=\omega^2+1$) which is readily recognized as the 
celebrated Aubry-Andr\'e-Harper model \cite{aubry1980analyticity,harper1955,dominguez2019aubry}.
A well-known feature of the model is that it displays a localization
transition at $\Delta=2$,  where the eigenmodes change from extended 
to exponentially localized. 
For a finite,  periodic  chain of length $N$ a typical 
choice is to set $\beta$ as a rational approximant, for
instance the ratio of consecutive Fibonacci numbers, 
such that for large $N$ $\beta$ approaches the golden mean.

\begin{figure}
\hfill\includegraphics[width=0.7\textwidth,clip]{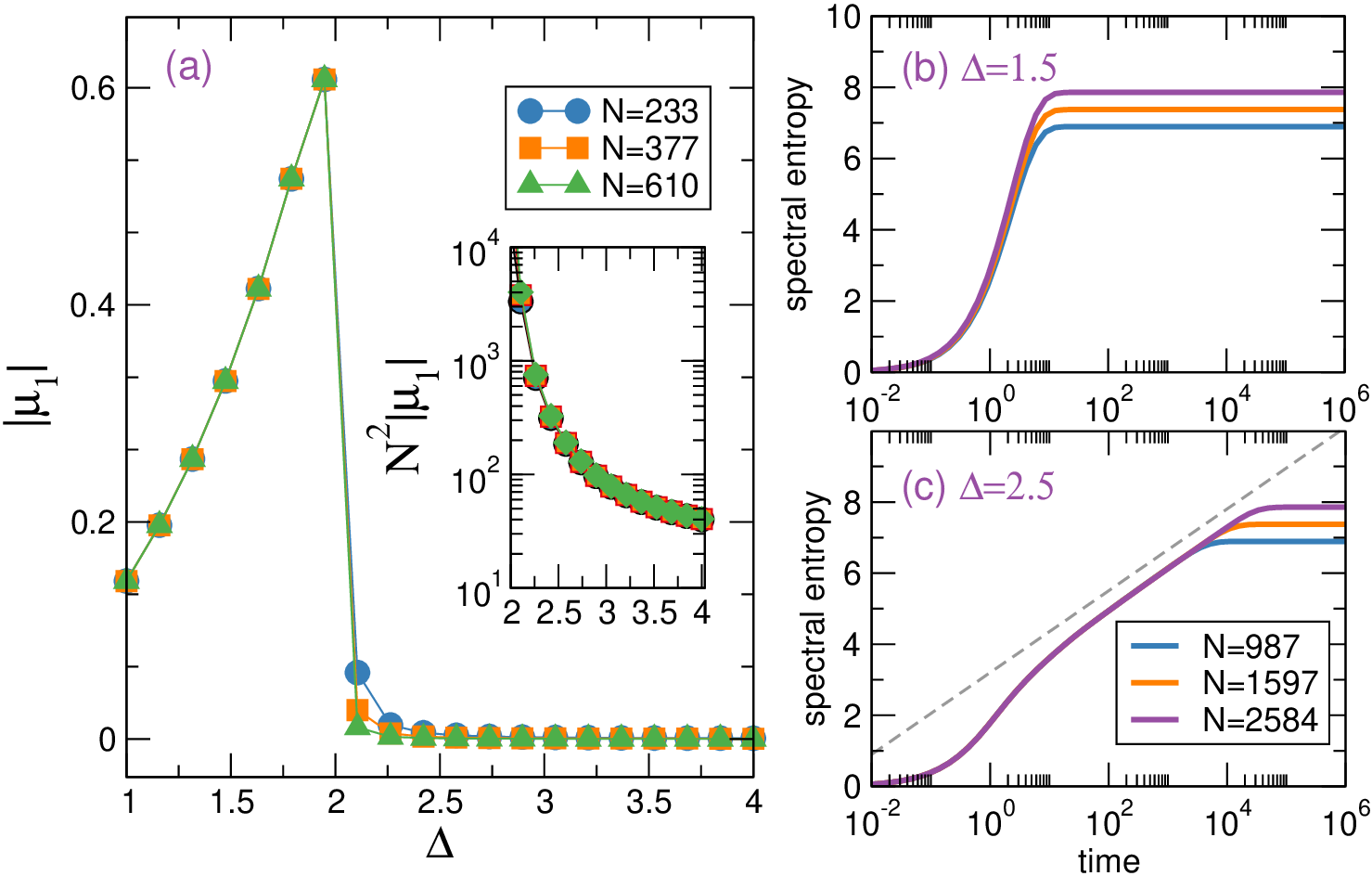}
\caption{Thermalization of the the quasi-periodic chain 
with with conservative noise
starting from the ground-state mode $\nu=1$, $\gamma=1$; 
(a) the spectral gap as a function of the potential strength $\Delta$, inset
shows that $\mu_2$ is proportional to $1/N^2$ for $\Delta>2$. (b,c)
time evolution of the spectral entropy $S(t)$ for two values $\Delta<2$ a
and $\Delta>2$ respectively. The dashed line is $\frac{1}{2}\log t$.
In all calculation we set $\beta$ as the ratio of consecutive Fibonacci numbers
and $\varphi=\pi/5$.}
\label{fig:gapaa}
\end{figure}

To compare with the disordered chain, in Figure  \ref{fig:gapaa}a
we report the spectral gap as a function of $\Delta$. 
For $\Delta<2$ the eigenvectors are extended and the spectral
gap remains finite and independent of the size.  Instead, 
as soon as localization 
sets in for $\Delta>2$,  $|\mu_1|$ vanishes as $1/N^2$ (see the 
inset of  Figure  \ref{fig:gapaa}a) and thermalization slows down considerably. 
The typical growth of the 
spectral entropy follow the same two basic routes as above:
$S(t)$ grows exponentially
or logarithmically in time for the two cases respectively
(see Figures  \ref{fig:gapaa}b-c) and saturates at the equipartition 
value.

Thus,  as far as average properties are concerned, the quasiperiodic 
lattice behaves very much as the disordered chain, the main difference
being that the 'closing'  of the spectral gap (and the change
from exponential to logarithmic growth of $S$) occurs sharply
at $\Delta=2$, independently of the lattice size.

As for the large deviation properties, we checked that upon increasing
the strength of the potential $\Delta$ the behavior is definitely similar to
the disordered chain.  For instance,  the qualitative behavior of $\lambda_+(s)$
for $\Delta=0.5$ is the same as that in Figures \ref{fig:topdis} (data 
not reported).  So we argue that the overall phase diagram should be 
similar in the two models, and schematically of the same form as in 
Figure (\ref{fig:pdiag}e) (with $w$ replaced by $\Delta$).

\begin{figure}
\hfill\includegraphics[width=0.8\textwidth]{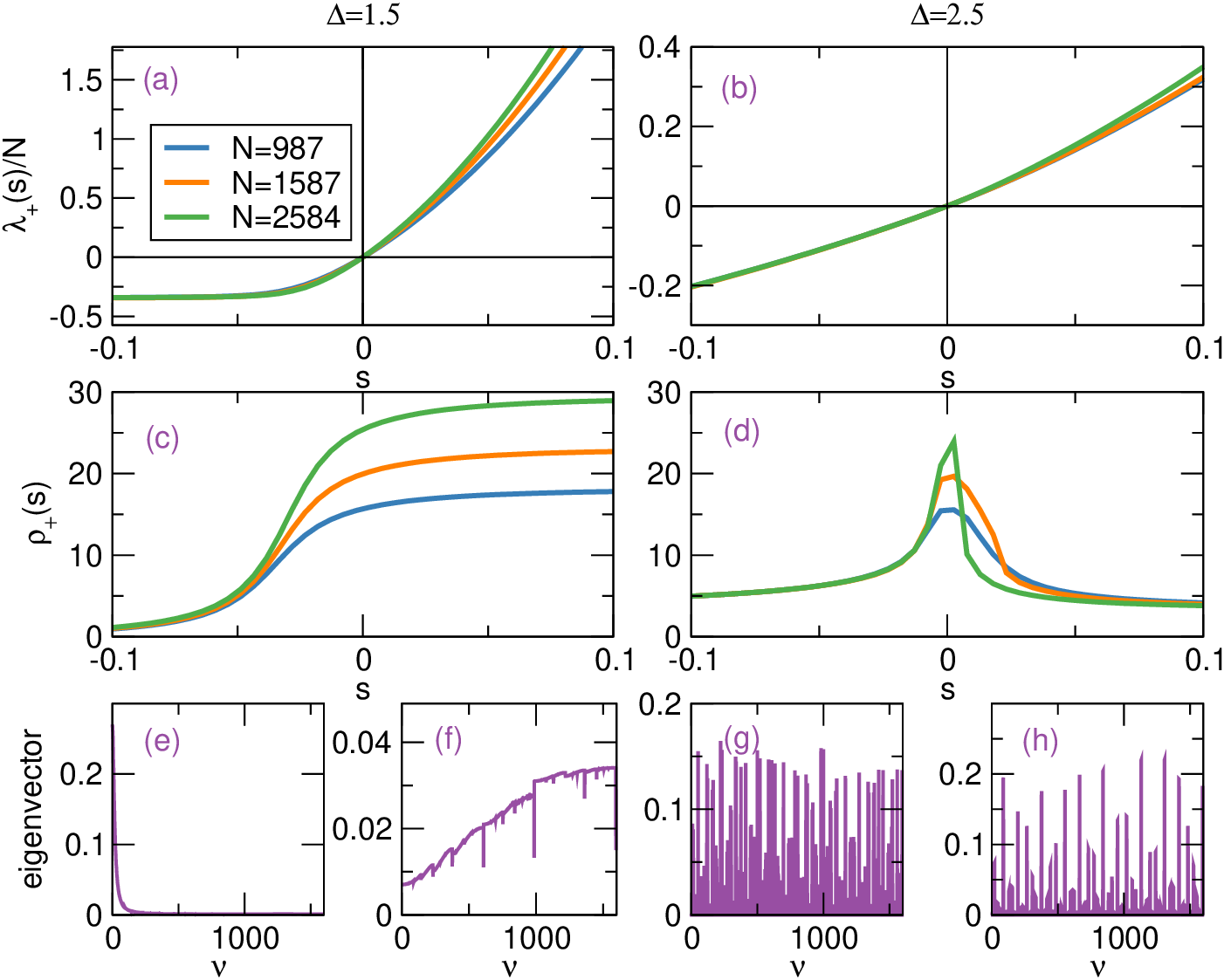}
\caption{Large deviations of dynamical entropy  for the 
quasiperiodic chain
with conservative noise $\gamma=1$ for different sizes and two value of the 
potential strength, $\Delta=1.5$ (left columns) and 
 $\Delta=2.5$ (right columns) ; (a,b) the scaled largest eigenvalue of $W_+(s)$ as 
a function of $s$  for different chain lengths $N$; 
(c,d) the order  parameter $\rho_+(s)$; (e-h) 
the eigenvectors $\phi_+$ for $N=1597$ as a function
of the lattice site number (e,g) $s=-0.1$ (f,h) $s=0.1$.}
\label{fig:topentaa}
\end{figure}

It is also of interest to study the behavior of the 
the large deviation functions across  the localization 
transition $\Delta=2$.  In particular the numerical solution shows that 
the dynamical phase transition does not occur around this value.
We just discuss the case of the entropy,  as the activity displays
similar features. 
For instance, in  Figure \ref{fig:topentaa}a,b  we report the 
scaled $\lambda_+$ for two values of $\Delta$ below and above
the transition. There is no sign of discontinuity neither in $\lambda_+$ 
nor in the order parameter $\rho_+$ ,
as shown in Figures \ref{fig:topentaa}c,d  which only shows 
a smooth monotonic growth in the extended phase $\Delta<2$ (Figure \ref{fig:topentaa}c) 
and a maximum 
around  $s=0$ in the localized phase $\Delta>2$ (Figure \ref{fig:topentaa}d).
Another difference is in the eigenvectors profiles that do not display
a sharp change from extended to localized as in the disordered chain
(compare Figure \ref{fig:topentaa}e-h with the lower panels of 
Figure \ref{fig:topdis}).  Actually, for $\Delta<2$, $\phi_+$ changes
gradually from a peaked 
distribution for negative $s$ to a uniform one for $s>0$, albeit with some 
structure (Figure \ref{fig:topentaa}e-f). Instead, for $\Delta>2$, 
$\phi_+$ remains extended 
and fluctuates wildly (Figure \ref{fig:topentaa}g-h), displaying a
fluctuating pattern
which may reflect the underlying structure of the eigenfunctions.

\section{Conclusions}

This work is an attempt to employ large-deviations methods to get 
insights on the thermalization of classical many-body systems. 
Exploiting the mathematical simplicity of 
conservative noise dynamics, we are able to compute the 
large deviation functions for two relevant quantities, 
the activity and dynamical entropy. For both observables,  
there is evidence  of a dynamical phase transitions
that correspond to trajectories of different qualitative nature.
We thus distinguish different region in action space 
corresponding to different thermalization pathways.

An appealing way to interpret the transition from extended to
localized in the leading eigenvectors, is to describe it 
as a form of condensation \cite{zannetti2014condensation,corberi2019probability} 
occurring in action/mode space. In other words,
most trajectories (a macroscopic fraction) in the localized phase concentrate on a 
subset of action space. The structure of the eigenvector
gives information on the actual modes mostly involved in the 
dynamics.  

When formulated in terms of  the kinetic equations, the thermalization of 
such system 
is basically a random walk on a large network in action space. Every 
move conserves the total energy and only changes the entropy
and the graph is undirected.  From this point of 
view,  there is a close connection 
with the general problem of large deviations of observables 
of random walks evolving on random graphs. This issue has been 
investigated recently \cite{debacco2016rare,coghi2019large}.
Indeed, sudden changes in degree fluctuations, similar to dynamical phase transitions, are related to localization transition are found that may be connected with 
what discussed here. 

\section*{Acknowledgements}

We thank Stefano Iubini for useful discussions and for reading the manuscript.

\section*{References}
\bibliographystyle{iopart-num}
\bibliography{heat.bib,disorder.bib,ldbib.bib}

\end{document}